\documentclass[aps,twocolumn,floatfix,showpacs,amssymb]{revtex4}
\usepackage{graphicx,bm}
\usepackage{hyperref}
\usepackage{amsmath}
\begin{document}
\title{Modified scalar product for the effective range approach:\\ the molecular contribution}
\author{Ludovic Pricoupenko}
\affiliation
{
Laboratoire de Physique Th\'{e}orique de la Mati\`{e}re Condens\'{e}e, 
Universit\'{e} Pierre et Marie Curie and CNRS, 4 place Jussieu, 75252 Paris, France.
}
\date{\today}
\begin{abstract}
The modified scalar product which permits to restore the self-adjoint character of the effective range approach is derived for one-channel contact models where a more general internal structure is included. In the case of the effective range approach, the modified scalar product  is interpreted in the light of a generic two-channel model for a narrow Feshbach resonance as a way to take into account implicitly the molecular contribution of the closed channel. 
\end{abstract}
%
\maketitle

The Wigner-Bethe-Peierls (WBP) zero-range model \cite{Wig33,Bet35} is very fruitful for obtaining low energy properties in the context of $s$ wave scattering resonances. In this regime, the scattering cross section is large with respect to its classical estimate ${\pi b^2}$ obtained from the knowledge of the characteristic range (or potential radius) $b$ of the pairwise interparticle potential. Consequently, for a $s$ wave resonance the scattering length (denoted $a$) is large in absolute value with respect to the potential radius. The WBP model is only parameterized by the scattering length and thus provides a minimal description of  systems in this resonant regime. In ultracold atoms, the resonance may be reached thanks to the Feshbach mechanism which involves a coherent coupling between different scattering channels, and also to the Zeeman effect which permits to change the molecular energies in the closed channels with respect to the continuum threshold in the open channel \cite{Koh06}. The resonance is then achieved by tuning an external magnetic field and this technique is usually called the magnetic Feshbach resonance. As a result, the scattering length $a$ may be tuned several orders of magnitude larger in absolute value than the typical potential radius of the pairwise interaction potential set by the van der Waals length
\cite{Gri93,Dal98}:
\begin{equation}
{R_{\rm vdW} =\left(\frac{\mu C_6}{\hbar^2}\right)^{1/4}} .
\label{eq:RvdW} 
\end{equation}
In Eq.~\eqref{eq:RvdW}, ${C_6}$ is the London's constant and $\mu$ is the reduced mass of the two colliding atoms. In the vicinity of the resonance the scattering length can be parameterized as \cite{Moe95}: 
\begin{equation}
a= a_{\rm bg} \left(1 - \frac{\Delta \mathcal B}{\mathcal B-\mathcal B_0} \right)
\label{eq:a(B)}
\end{equation}
where ${a_{\rm bg}}$ is the background scattering length and ${\Delta \mathcal B}$ is the width of the resonance. Equation \eqref{eq:a(B)} is derived from a two-channel description of the resonance mechanism which permits also to describe quantitatively two-body scattering properties at low energies. As a result, two-channel models provide a characteristic length related to the width of the magnetic Feshbach resonance and  referred in what follows as the width radius \cite{Pet04b,Koh06}:
\begin{equation}
R^\star= \frac{\hbar^2}{m a_{\rm bg} \delta \mu \Delta \mathcal B } .
\label{eq:Rstar}
\end{equation}
For broad resonances, the effective range is of the order of the van der Waals length and the WBP model is well adapted for the description of the interaction in this regime. However in the case of narrow resonances, the width radius has an anomalously large value with respect to ${R_{\rm vdW}}$ and the effective range  ${r_{\rm e} \sim -2 R^\star}$ is large and negative. In this last resonant regime, it is also possible to model the interaction by using the effective range approach where the zero-range force is parameterized by the scattering length and the width radius \cite{Pet04b}. More generally, it is sometimes relevant to consider a one-channel zero-range model which reproduces the expression of the scattering phase-shift at low energy and thus somehow encapsulates the internal structure of the interaction \cite{Bold03,Nai07,Pri10b}. Such generalized WBP model lead to what can be called a non standard quantum mechanics in the sense that the Hamiltonian of the system is not self-adjoint. 

In Ref.~\cite{Pri06a} a modified scalar product was introduced in order to restore the self-adjoint character of the effective range approach. For a bound state, the modified scalar product possesses the interesting property of giving the same normalization constant as the one supported by the analyticity of the scattering amplitude \cite{Lan99}. In section \ref{sec:prodscal} of this paper, the modified scalar product is generalized for a large class of zero-range models having an internal structure. In section \ref{sec:2channel} a generic two-channel model is introduced for describing narrow resonances. Finally in section \ref{sec:fusion}, the expression of the modified scalar product introduced for the effective range approach is derived from the two-channel model. This permits to realize that the metrics associated to the effective range approach is a way to take implicitly into account the closed channel contribution in a zero-range one-channel model.

\section{Zero-range potential approach}
\label{sec:prodscal}
\subsection{Contact condition}

The WBP model can be introduced without loss of generality by considering two spinless interacting particles of respective  positions ${\mathbf r_1}$, ${\mathbf r_2}$ and  reduced mass $\mu$. In this model, the wave function ${\Psi(\mathbf r_1,\mathbf r_2)}$ of this system verifies the free (\emph{i.e.} non interacting) Schr\"{o}dinger equation for all values of the  relative coordinates ${\mathbf r=\mathbf r_1-\mathbf r_2}$ excepted at the contact (${\mathbf r=0}$). Moreover at a fixed position of the center of mass of the two particles, the interaction is replaced by an asymptotic condition on the wavefunction for vanishing relatives coordinates ${\mathbf r \to 0}$:
\begin{equation}
\Psi(\mathbf r_1,\mathbf r_2) =  A_\Psi \times \left(\frac{1}{a}-\frac{1}{r} \right) + O(r) ,
\label{eq:contactWBP}
\end{equation}
where $a$ is the $s$ wave scattering length and ${A_\Psi}$ is not a function of $r$. Equation~\eqref{eq:contactWBP} is often called the contact condition of the WBP model. The regime of validity of the WBP model is limited to situations where the relative energy of the two particles ${E_{\rm rel}}$ (\emph{i.e.} the energy of the two colliding particles in their center of mass frame) is much less in absolute value than ${\hbar^2/(\mu r_{\rm e}^2)}$, where ${r_{\rm e}}$ is the effective range of the actual finite-range interparticle potential. The condact condition in Eq.~\eqref{eq:contactWBP} has been generalized in various studies where the inverse scattering length ${1/a}$ is substitued by a function ${-g(E)}$ which depends on the relative energy ${E=\frac{\hbar^2k^2}{m}}$, in such a way that the zero-range approach can support an arbitrary $s$ wave scattering amplitude:
\begin{equation}
f_s(k)=\frac{-1}{-g(E)+ik} .
\label{eq:f(k)}
\end{equation}
In Eq.~\eqref{eq:f(k)} the function ${g(E)}$ is defined from the scattering $s$ wave phase shift ${\delta_s(k)}$ by
\begin{equation}
g(E) =  k \cot \delta_s(k) .
\label{eq:g(E)}
\end{equation}
However, this  method which appears to be powerful for obtaining two-body properties in inhomogeneous situations \cite{Bold03,Nai07}, is not satisfactory for relative energies of the order of (or greater than) the van der Waals radius ${E_{\rm vdW}}$ where higher partial waves have non negligible contributions to the scattering process.

In the vicinity of a narrow magnetic Feshbach resonance, the width radius in Eq.~\eqref{eq:Rstar} is large with respect to ${R_{\rm vdW}}$ and the low energy behavior of the scattering amplitude is consistently parameterized in the region of asymptotically small magnetic detunings
${|{(\mathcal B - \mathcal B_0)} / {\Delta \mathcal B}| \ll 1}$ by using the expression \cite{Pet04b,Chi10,Pri11c}: 
\begin{equation}
g(E)=-\frac{1}{a} - R^\star  k^2 .
\label{eq:approx_re}
\end{equation}
In Refs.~\cite{Pri08,Pri11a,Pri11b}, it has been shown that it can be fruitful to relax the strict zero-range limit in intermediary calculations while keeping the structure of the WBP model \cite{uniform}. In the subsequent lines, this technique is briefly introduced by considering the two-body system in the center of mass frame without any external field. In this approach, the actual interaction in the Schr\"{o}dinger equation is replaced by a source term of vanishing range. An eigenstate  ${| \Psi_\epsilon\rangle}$ (energy ${E_\Psi}$) of the stationnary Schr\"{o}dinger equation satisfies:
\begin{equation}
\left(\frac{\mathbf p^2}{2\mu} - E_\Psi \right)  | \Psi_\epsilon \rangle =  - \frac{2 \pi \hbar^2 A_\Psi}{\mu} | \delta_\epsilon \rangle ,
\label{eq:SchrodEpsi}
\end{equation}
where ${\langle \mathbf r | \delta_\epsilon \rangle}$ converges toward the $\delta$ distribution in the zero-range limit where ${\epsilon}$ tends to zero. In Eq.~\eqref{eq:SchrodEpsi}, ${\mathbf p}$ is the momentum operator for the relative particle and the source amplitude ${A_\Psi}$ is given by
\begin{equation}
 A_\Psi = \frac{-1}{4\pi} \times \lim_{k\to \infty} k^2 \lim_{\epsilon \to 0} \langle \mathbf k |\Psi_\epsilon \rangle .
\label{eq:APsi}
\end{equation}
A natural choice for ${\langle \mathbf r |\delta_\epsilon\rangle}$ is to consider a family of Gaussian functions parameterized by $\epsilon$:
\begin{equation}
\langle {\mathbf r} | \delta_\epsilon \rangle = \frac{1}{(2 \pi \epsilon^2)^{3/2}} \exp\left(-\frac{r^2}{\epsilon^2}\right) .
\end{equation}
These functions have a simple expression in the momentum representation: ${\langle {\mathbf k} | \delta_\epsilon \rangle=\chi_\epsilon(k)}$ where
\begin{equation}
\chi_\epsilon(k) = \exp(-k^2 \epsilon^2/4) .
\label{eq:chi}
\end{equation}
In Eq.~\eqref{eq:chi} the ket ${|\mathbf k \rangle}$ is choosen with the usual convention of scattering theory: ${\langle r | \mathbf k  \rangle=\exp(i\mathbf k \cdot \mathbf r)}$. In this framework,  a possible form of the generalized WBP contact condition can be written as \cite{Pri11a}:
\begin{equation}
\lim_{\epsilon \to 0} \partial_\epsilon \left[ \epsilon \langle \delta_\epsilon | \Psi_\epsilon \rangle \right] = - A_\Psi g(E_\Psi).
\label{eq:Regepsilon}
\end{equation}
The zero range generalized WBP model is exactly recovered by performing the limit where $\epsilon$ tends to zero at the end of the calculation. In what follows, the index ${\epsilon}$ is dropped when the zero-range limit is achieved. For instance: ${|\Psi \rangle = \lim_{\epsilon \to 0} |\Psi_\epsilon \rangle}$.

\subsection{The modified scalar product}

Generalized WBP models where ${g(E)}$ in Eq.~\eqref{eq:g(E)} is not a constant, are not self-adjoint with respect to the usual scalar product. This problem has been solved for the effective range approach in Ref.~\cite{Pri06a} by introducing a modified scalar product. In what follows, the modified scalar product is derived for a general form of the function ${g(E)}$. For this purpose, one considers two eigenstates denoted by ${|\Phi_\epsilon \rangle}$ and ${|\Psi_\epsilon \rangle}$, where  ${|\Psi_\epsilon \rangle}$ verifies Eqs.~(\ref{eq:SchrodEpsi},\ref{eq:Regepsilon}) and ${|\Phi_\epsilon \rangle}$ verifies:
\begin{eqnarray}
\left(\frac{\mathbf p^2}{2\mu} - E_\Phi \right) |\Phi_\epsilon \rangle = -\frac{2\pi\hbar^2 A_\Phi}{\mu} |\delta_\epsilon\rangle
\label{eq:SchrodEphi}
\end{eqnarray}
and also the contact condition analogous to Eq.~\eqref{eq:Regepsilon}. For ${\epsilon \ne 0}$, the functions ${\langle \mathbf r |\Psi_\epsilon \rangle}$ and ${\langle \mathbf r |\Psi_\epsilon \rangle}$ are regular at the contact ${\mathbf r=0}$. Thus the identity ${\langle \Phi_\epsilon | p^2 |\Psi_\epsilon \rangle = \langle \Psi_\epsilon | p^2 |\Phi_\epsilon \rangle^*}$ holds and for two nondegenerate states ${(E_\Phi \ne E_\Psi)}$, from Eqs.~(\ref{eq:SchrodEpsi},\ref{eq:SchrodEphi}) one obtains the following expression for the usual scalar product:
\begin{equation}
\langle \Phi_\epsilon |\Psi_\epsilon \rangle = \frac{2 \pi \hbar^2}{\mu} \times \frac{ A_{\Phi}^* \langle \delta_\epsilon |\Psi_\epsilon \rangle- A_\Psi \langle \Phi_\epsilon |\delta_\epsilon \rangle}
{E_\Phi -E_\Psi} . 
\label{eq:scal1}
\end{equation}
The regular part of  ${\langle \delta_\epsilon |\Psi_\epsilon \rangle}$ in Eq.~\eqref{eq:scal1} is deduced straightforwardly from the contact condition in Eq.~\eqref{eq:Regepsilon} while its singular ${1/\epsilon}$ behavior in the zero-range limit ${(\epsilon\to 0)}$ follows from Eq.~\eqref{eq:SchrodEpsi}. Finally, in the zero-range limit ${(\epsilon \to 0)}$ one obtains:
\begin{equation}
\langle \delta_\epsilon |\Psi_\epsilon \rangle \operatornamewithlimits{=}_{\epsilon \to 0} 
-A_\Psi \left[ \frac{1}{\epsilon} \sqrt{\frac{2}{\pi}}  + g(E_\Psi) \right] + {\mathcal O}(\epsilon) .
\end{equation}
Using this last relation and its analog for ${\langle \delta_\epsilon |\Phi_\epsilon \rangle}$, the usual scalar product in Eq.~\eqref{eq:scal1} can be written in the zero-range limit as:
\begin{equation}
\langle \Phi |\Psi \rangle = \frac{2 \pi \hbar^2A_{\Phi}^*A_\Psi}{\mu} \times  \frac{g(E_\Phi)-g(E_\Psi)}{E_\Phi -E_\Psi}  .
\label{eq:prod_scal} 
\end{equation}
The modified scalar product denoted by ${(\Phi |\Psi )_0}$ is such that two non degenerate states are mutually orthogonal. It is thus built by substracting the right hand side of Eq.~\eqref{eq:prod_scal} to the usual scalar product ${\langle \Phi |\Psi \rangle}$. 

If a bound state exists ${|\phi\rangle}$, its (negative) energy  ${E_\phi=-\hbar^2q^2/m}$ is a pole of Eq.~\eqref{eq:f(k)} where ${k=iq}$ and ${q>0}$. In the zero-range model, the wave function of this bound state is ${\langle \mathbf r | \phi\rangle = - A_\phi \exp(-qr)/r}$, \emph{i.e.} it corresponds to the asymptotic form for large interparticle distances $({r\to \infty)}$ of an actual bound state. By imposing that the state is normalized to unity with respect to the modified scalar product [\emph{i.e.} ${(\phi |\phi )_0=1}$], one obtains:
\begin{equation}
|A_\phi|^2 = \frac{1}{4\pi} \times \frac{2}{\frac{1}{q}-\frac{\hbar^2}{\mu} g'(E_\phi)} .
\label{eq:normal_bound}
\end{equation}
The result of Eq.~\eqref{eq:normal_bound} exactly coincides with the one derived by using the analyticity of the scattering amplitude~\cite{Lan99}.

For the effective range approach, from Eqs.~(\ref{eq:approx_re},\ref{eq:prod_scal}) the modified scalar product can be written as:
\begin{equation}
(\Phi |\Psi)_0 = \int d^3r \left[ 1+ R^\star \delta(r)\right] \Phi^*(\mathbf r) \Psi(\mathbf r) .
\label{eq:prod_scal_modif}
\end{equation}
This last expression may be generalized if one assumes that the function ${g(E)}$ admits an expansion in powers of $E$ in the vicinity of zero energy:
\begin{equation}
g(E) = \sum_{n=0}^\infty c_n E^n ,
\end{equation}
which is true if one considers that the present zero-range approach models a pairwise short-range potential. For this purpose, one introduces the regular part of the following limit valid for all non-zero positive integer $n$:
\begin{equation}
\operatornamewithlimits{Reg}_{k\to \infty} \langle \mathbf k | (\frac{\mathbf p^2}{2\mu})^n |\Psi \rangle = - \frac{2\pi \hbar^2 A_\Psi}{\mu} \times E_\Psi^{n-1} 
\end{equation}
while this regularized limit is set to zero for ${n=0}$. Using this tool, the modified scalar product can be finally expressed as:
\begin{multline}
(\Phi|\Psi)_0 = \langle \Phi | \Psi \rangle -\frac{\mu}{2\pi \hbar^2} \sum_{n=1}^\infty c_n \sum_{p=1}^n
\operatornamewithlimits{Reg}_{k\to \infty} \langle \Phi | (\frac{\mathbf p^2}{2\mu})^{n-p+1} |\mathbf k \rangle \\
\times \operatornamewithlimits{Reg}_{k\to \infty} \langle \mathbf k | (\frac{\mathbf p^2}{2\mu})^{p} |\Psi \rangle .
\end{multline}

\section{The generic two-channel model for a narrow resonance}

\label{sec:2channel}

The generic two-channel model used in this paper is similar to the ones introduced in Refs.~\cite{Jon08,Gog08} and is a simplified form of the models of Refs.~\cite{Koh06, Wer09,Jon10}. The simplest way to introduce the model is to use the second quantization in the momentum representation. In what follows only the case of a bosonic neutral spinless particles is considered, nevertheless similar equations and same conclusions can be drawn  about the modified scalar product if one considers fermions or heteronuclear systems. Bosons in the open channel have a mass denoted by $m$ and the threshold for the continuum states is fixed at zero energy. The operator ${a_{\mathbf k}}$ (respectively  ${b_{\mathbf k}}$) creates an atomic (respectively a molecular) plane wave  ${\langle \mathbf r|{\mathbf k} \rangle =  \exp({i {\mathbf k} \cdot {\mathbf r}})}$. The creation and annihilation operators obey the usual commutation rules:
\begin{equation}
[a_{\mathbf k}, a_{\mathbf k'}^\dagger]=(2\pi)^3 \delta(\mathbf k-\mathbf k') \ , \  [b_{\mathbf k}, b_{\mathbf k'}^\dagger]=(2\pi)^3 \delta(\mathbf k-\mathbf k'), 
\label{eq:commut}
\end{equation}
and the others possible commutators vanish. The Feshbach resonance mechanism is encapsulated by a coherent coupling between atomic pairs of the open channel and the molecular state. In what follows, the kinetic energy of an atomic plane wave of wave vector ${\mathbf k}$ is 
\begin{equation}
\epsilon_{\mathbf k} = \frac{\hbar^2 k^2}{2m} .
\label{eq:epsilon_k}
\end{equation}
The Hamiltonian of the generic two-channel model used in this paper is:
\begin{multline}
H =\int \frac{d^3k}{(2\pi)^3} \epsilon_{\mathbf{k}}  a^\dagger_{\mathbf k} a_{\mathbf k} + \int \frac{d^3K}{(2\pi)^3} 
\left( \frac{ \epsilon_{\mathbf K}}{2} +E_{\rm mol} \right) b^\dagger_{\mathbf K} b_{\mathbf K} \\
+ \Lambda \left[ \int 
\frac{d^3Kd^3k}{(2\pi)^6}  \, \chi_\epsilon(\mathbf k) \, b^\dagger_{\bf K} a_{\frac{{\mathbf K} }{2}-{\mathbf k}} a_{\frac{{\mathbf K}}{2}+{\mathbf k}} + {\rm h. c.}  \right] .
\label{eq:H}
\end{multline}
In Eq.~\eqref{eq:H}, ${E_{\rm mol}}$ is the internal energy of the molecular state and the last term couples the two channels and thus models the Feshbach resonance mechanism. For simplicity the interchannel coupling is described by the generic function ${\chi_\epsilon}$ of Eq.~\eqref{eq:chi} with an amplitude denoted by $\Lambda$. 

\section{Interpretation of the modified scalar product}

\label{sec:fusion}

In this part, the modified scalar product in Eq.~\eqref{eq:prod_scal_modif} associated to the effective range approach is derived from the zero-range limit of the two-channel model. In this model a two-body state is a coherent superposition of two atoms in the open-channel and one molecule in the closed-channel. In the center of mass frame, it can be written as
\begin{equation} 
\label{2b-psi-ansatz}
|\Psi_{\rm tot} \rangle = |\Psi_{\rm open}\rangle  + \beta_\Psi b^\dagger_{\mathbf 0} |0 \rangle  
\end{equation}
where ${|\Psi_{\rm open}\rangle}$ is the atomic state
\begin{equation}
|\Psi_{\rm open}\rangle = \int \frac{d^3k}{(2\pi)^3} \alpha_\Psi(\mathbf k) a^\dagger_{\mathbf k} a^\dagger_{-\mathbf k} | 0 \rangle 
\end{equation}
characterized in the momentum representation by the amplitude  ${\alpha_\Psi(\mathbf k)}$ and ${\beta_\Psi}$ is the amplitude associated to the molecular part. The usual scalar product between two states ${|\Phi_{\rm tot}\rangle}$ and ${|\Psi_{\rm tot}\rangle}$ is given by
\begin{equation}
\langle \Phi_{\rm tot} |\Psi_{\rm tot} \rangle = \langle \Phi_{\rm open} |\Psi_{\rm open} \rangle + \beta_\Phi^* \beta_\Psi .
\label{eq:vrai_prod}
\end{equation}
Projection of the stationnary Schr\"{o}dinger equation in the open- and closed-channel gives the following system of equations:
\begin{eqnarray} 
&&\left( E_\Psi - 2 \epsilon_{\mathbf k } \right) \alpha_\Psi(\mathbf k)-\Lambda \beta_\Psi \chi_\epsilon(\mathbf k) = 0 \label{eq:schro-alpha} \\ 
&&(E_\Psi -E_{\rm mol}) \beta_\Psi - 2 \Lambda  \int \frac{d^3 k}{(2\pi)^3} \chi_\epsilon(\mathbf k) \alpha_\Psi(\mathbf k) = 0 .
\label{eq:schro-beta}
\end{eqnarray}
For a scattering state of incoming wave vector ${\mathbf k_0}$, ${E_\Psi=\hbar^2k_0^2/(2m)}$ and the atomic amplitude is:
\begin{equation}
 \alpha_\Psi(\mathbf k) = (2\pi)^3 \delta(\mathbf k - \mathbf k_0) + \frac{\Lambda \chi_\epsilon(\mathbf k) \beta_\Psi}{E_\Psi - 2\epsilon_{\mathbf k}+i 0^+} .
\end{equation}
The atomic scattering amplitude is thus given by 
\begin{equation}
f(E_\Psi) =-\frac{\mu}{2\pi \hbar^2} \Lambda \chi_\epsilon(\mathbf k) \beta_\Psi .
\label{eq:f2channelmodel}
\end{equation}
From Eqs.~(\ref{eq:schro-alpha},\ref{eq:schro-beta},\ref{eq:f2channelmodel}) one obtains:
\begin{equation}
f(E_\Psi) = - \frac{m}{4\pi\hbar^2} \frac{[\chi_\epsilon(k_0)]^2}
{
\frac{(E_\Psi-E_{\rm mol})}{2 \Lambda^2} -
\int \frac{d^3k}{(2\pi)^3} \frac{[\chi_\epsilon(k)]^2}{E_\Psi-\frac{\hbar^2 k^2}{m} +i0^+}
} .
\label{eq:scat-amplitude}
\end{equation}
The scattering length $a_\epsilon$ of the model is given by ${-1/f(0)}$ and the parameter $R^\star_\epsilon$, by the term in $k^2$ in the low energy expansion of ${1/f(E_\Psi)}$. One finds:
\begin{align} 
\frac{1}{a_\epsilon} &= \frac{1}{\epsilon} \sqrt{\frac{2}{\pi}} - \frac{2 \pi \hbar^2 E_{\rm mol}}{\Lambda^2 m} \\
R^\star_\epsilon &= - \sqrt{\frac{2}{\pi}} \epsilon + \frac{2\pi\hbar^4}{\Lambda^2 m^2} + \frac{\epsilon^2}{2a} .
\label{eq:a&re}
\end{align}
In the zero-range limit (${\epsilon\to0}$) the parameters $a$ and $R^\star$ can be kept fixed at a given value (in this formal limit, the molecular energy ${E_{\rm mol}}$ tends to infinity), and one finds a scattering amplitude which is exactly the one of the effective range approach where 
\begin{equation}
R^\star=\frac{2\pi \hbar^4}{m^2\Lambda^2} .
\label{eq:re_Lambda}
\end{equation}
In the momentum representation, the wave function in the open channel is:
\begin{equation}
\langle {\mathbf k} | \Psi_{\rm open} \rangle = \frac{\alpha_\Psi(\mathbf k) +  \alpha_\Psi(-\mathbf k)}{\sqrt{2}} . 
\label{eq:Psiopen}
\end{equation}
In the mapping between the effective range approach and the two-channel model, in the limit where $\epsilon$ tends to zero, ${| \Psi_{\rm open} \rangle}$ coincides with the state ${| \Psi_\epsilon \rangle}$ in Eq.~\eqref{eq:SchrodEpsi}. Thus, using the definition of ${A_\Psi}$ in Eq.~\eqref{eq:APsi} in the limit where ${\epsilon}$ tends to zero, Eqs.~(\ref{eq:schro-alpha},\ref{eq:Psiopen}) provide the relation:
\begin{equation}
A_{\Psi_{\rm open}}=\frac{\sqrt{2}m\Lambda\beta_\Psi}{4\pi\hbar^2} .
\label{eq:lien}
\end{equation}
From Eq.~(\ref{eq:vrai_prod},\ref{eq:re_Lambda},\ref{eq:lien}) one finally obtains:
\begin{equation}
\langle \Phi_{\rm tot} |\Psi_{\rm tot} \rangle = \langle \Phi_{\rm open} |\Psi_{\rm open} \rangle + 4 \pi R^\star A_\Phi^* A_\Psi .
\label{eq:prodscal-2voies}
\end{equation}
The bit ${\langle \Phi_{\rm open} |\Psi_{\rm open} \rangle}$ in Eq.~\eqref{eq:prodscal-2voies} corresponds to the usual scalar product ${\langle \Phi |\Psi \rangle}$ in Eq.~\eqref{eq:prod_scal_modif} and by simple identification ${\langle \Phi_{\rm tot} |\Psi_{\rm tot} \rangle}$ coincides with the modified scalar product ${(\Phi|\Psi)_0}$ in Eq.~\eqref{eq:prod_scal_modif}. Hence, Eq.~\eqref{eq:prodscal-2voies} shows that the modified scalar product introduced in the effective range approach is nothing but a way to take into account implicitly the contribution of the closed channel in a single-channel zero-range theory.

\section{Conclusions}

In this paper, the modified scalar product initially introduced for the mathematical consistency of the effective range approach is depicted in terms of the contribution of a hiden closed channel. The modified scalar product is also generalized for a zero-range model built from the expansion of the $s$-wave scattering phase shift in integer powers of the collisional energy. More generally, this paper illustrates the fact that the modified scalar product is a way to take into account the shot range contribution of the wavefunction in a formal one-channel zero-range model.

\section*{Acknowledgments}

Y. Castin and M. Jona-Lasinio are acknowledged for discussions. The Laboratoire de Physique Th\'{e}orique de la Mati\`{e}re Condens\'{e}e is Unit\'{e} Mixte de Recherche 7600 of Centre National de la Recherche Scientifique and its Cold Atoms group is associated with Institut  Francilien de Recherche sur les Atomes Froids.


\begin{thebibliography}{0}
\expandafter\ifx\csname natexlab\endcsname\relax\def\natexlab#1{#1}\fi
\expandafter\ifx\csname bibnamefont\endcsname\relax
  \def\bibnamefont#1{#1}\fi
\expandafter\ifx\csname bibfnamefont\endcsname\relax
  \def\bibfnamefont#1{#1}\fi
\expandafter\ifx\csname citenamefont\endcsname\relax
  \def\citenamefont#1{#1}\fi
\expandafter\ifx\csname url\endcsname\relax
  \def\url#1{\texttt{#1}}\fi
\expandafter\ifx\csname urlprefix\endcsname\relax\def\urlprefix{URL }\fi
\providecommand{\bibinfo}[2]{#2}
\providecommand{\eprint}[2][]{\url{#2}}

\end{thebibliography}


\begin{thebibliography}{99}

\bibitem{Wig33} E. Wigner,  Zeits. f. Physik {\bf 83}, 253 (1933).

\bibitem{Bet35} H. Bethe and R. Peierls,  Proc. R. Soc. London, Ser. A {\bf 148}, 146 (1935).

\bibitem{Koh06} T. K\"{o}hler, K. G\'{o}ral, and P.S. Julienne, 
\href{http://dx.doi.org/10.1103/RevModPhys.78.1311}
{Rev. Mod. Phys. {\bf 78}, 1311 (2006)}.

\bibitem{Gri93} G.F. Gribakin and V.V. Flambaum, 
\href{http://dx.doi.org/10.1103/PhysRevA.48.546}
{Phys. Rev. A {\bf 48}, 546 (1993)}.

\bibitem{Dal98} J. Dalibard, Proceedings of the International School of  Physics 'Enrico  Fermi', Course CXL: 
{\sl 'Bose-Einstein condensation in gases'}, Varenna 1998,  M. Inguscio, S. Stringari, C. Wieman edts.

\bibitem{Moe95} A. J. Moerdijk, B. J. Verhaar, and A. Axelsson, 
\href{http://dx.doi.org/10.1103/PhysRevA.51.4852}
{Phys. Rev. A {\bf 51}, 4852 (1995)}.

\bibitem{Pet04b} D.S. Petrov, 
\href{http://dx.doi.org/10.1103/PhysRevLett.93.143201}
{Phys. Rev. Lett. {\bf 93}, 143201 (2004)}.

\bibitem{Bold03} E. L. Bolda, E. Tiesinga, and P. S. Julienne, 
\href{http://dx.doi.org/10.1103/PhysRevA.68.032702}
{Phys. Rev. A {\bf 68}, 032702 (2003)}.

\bibitem{Nai07} P. Naidon, E. Tiesinga, W.F. Mitchell, P.S. Julienne, 
\href{http://dx.doi.org/10.1088/1367-2630/9/1/019}
{New J. Phys. {\bf 9}, 19 (2007)}. 

\bibitem{Pri10b} L. Pricoupenko, 
\href{http://dx.doi.org/10.1103/PhysRevA.82.043633}
{Phys. Rev. A {\bf 82}, 043633 (2010)}.

\bibitem{Pri06a} L. Pricoupenko, 
\href{http://dx.doi.org/10.1103/PhysRevA.73.012701}
{Phys. Rev. A {\bf 73}, 012701 (2006)}.

\bibitem{Lan99} L. Landau and E. Lifchitz, 'Quantum Mechanics' (Butterworth-Heinemann, Oxford, 1999).

\bibitem{Chi10} C. Chin, R. Grimm, P. Julienne, and E. Tiesinga, 
\href{http://dx.doi.org/10.1103/RevModPhys.82.1225}
{Rev. Mod. Phys. {\bf 82}, 1225 (2010)}.

\bibitem{Pri11c} L. Pricoupenko, M. Jona-Lasinio, {\sl arXiv:1109.3002}.

\bibitem{Pri08} L. Pricoupenko, 
\href{http://dx.doi.org/10.1103/PhysRevLett.100.170404}
{Physical Review Letters {\bf 100}, 170404 (2008)}.

\bibitem{Pri11a} L. Pricoupenko, 
\href{http://dx.doi.org/10.1103/PhysRevA.83.062711}
{Phys. Rev. A {\bf 83}, 062711 (2011)}.

\bibitem{Pri11b} L. Pricoupenko, to be published in Phys. Rev. A (2011).

\bibitem{uniform} In the strict zero-range limit the function ${r \langle \mathbf r |\Psi\rangle}$ where  ${ \langle \mathbf r |\Psi\rangle}$ satisfies Eq.~\eqref{eq:contactWBP}, has a non zero limit at the contact ${\mathbf r=0}$ while if one considers the expansion of ${|\Psi\rangle}$ on a basis of regular functions, each term of the series for ${r \langle \mathbf r |\Psi\rangle}$ is zero. Consequently, the limit of the series expansion and the zero range limit do not commutes: there is no uniform convergence. This property is discussed in depth in the lecture 4 of Claude Cohen Tannoudji at the  Coll\`{e}ge de France (year 1998-1999) \cite{CCT}. The problem of commutation of the two limits is avoided by substituting the delta distribution in the zero range force by a function of arbitrarily small (but finite) range (see the right hand side of Eq.~\eqref{eq:SchrodEpsi}) together with the contact condition of Eq.~\eqref{eq:Regepsilon}.  
%
\bibitem{CCT} C. Cohen-Tannoudji, {'Cours du Coll\`{e}ge de France'} (1998-99);
\href{http://www.phys.ens.fr/cours/college-de-france/}{http://www.phys.ens.fr/cours/college-de-france/}.

\bibitem{Jon08}  M. Jona-Lasinio, L. Pricoupenko and Y. Castin, 
\href{http://dx.doi.org/10.1103/PhysRevA.77.043611}
{Phys. Rev. A {\bf 77}, 043611 (2008)}.

\bibitem{Gog08} A.O. Gogolin, C. Mora, R. Egger, 
\href{http://dx.doi.org/10.1103/PhysRevLett.100.140404}
{Phys. Rev. Lett. {\bf 100}, 140404 (2008)}.

\bibitem{Wer09} F. Werner, L. Tarruell, and Y. Castin, 
\href{http://dx.doi.org/10.1140/epjb/e2009-00040-8}
{Eur. Phys. J. B {\bf 68}, 401 (2009)}.

\bibitem{Jon10} M. Jona-Lasinio, L. Pricoupenko, 
\href{http://dx.doi.org/10.1103/PhysRevLett.104.023201}
{Phys. Rev. Lett. {\bf 104}, 023201 (2010)}.

\end{thebibliography}
\end{document}